\documentclass[smallextended]{svjour3}       
\smartqed 
\usepackage{graphicx}
\graphicspath{{images/}}
 \usepackage{mathptmx}   
 \usepackage{caption}   

\journalname{J Low Temp Phys
	 DOI 10.1007/s10909-016-1483-2}

\begin{document}

\title{Analysis of electronic  and structural properties of surfaces and interfaces based on LaAlO$ _{3} $ and SrTiO$ _{3}$
}

\author{Piyanzina~I.I.  \and Lysogorskiy~Yu.V.      \and Varlamova~I.I.  \and Kiiamov~A.G. \and  Kopp~T. \and Eyert~V. \and Nedopekin~O.V. \and Tayurskii~D.A.
}

\institute{Piyanzina~I.I., Lysogorskiy~Yu.V., Varlamova~I.I., Kiiamov~A.G., Nedopekin O.V., Tayurskii~D.A. 
		\at Institute of Physics, Kazan Federal University, Kremlyovskaya St.~18, 420008 Kazan, Russia \\
              Piyanzina~I.I., Kopp~T.
              \at EP VI and Center for Electronic Correlations and Magnetism, Universit\"at Augsburg, Universit\"atsstra{\ss}e~1, D-86135 Augsburg, Germany \\
              \email{irina.piyanzina@physik.uni-augsburg.de \\
              Eyert~V. 
              \at Materials Design SARL, 18 rue de Saisset, 92120 Montrouge, 
              France\\
              }  
}

\date{Received: 4 June 2015 / Accepted: 4 January 2015}

\maketitle

\begin{abstract}
Recently, it was established that a two-dimensional electron system can arise at the interface between two oxide insulators LaAlO$_{3}$ and SrTiO$_{3}$. This paradigmatic example exhibits metallic behaviours and magnetic properties between non-magnetic and insulating oxides. Despite a huge amount of theoretical and experimental work a thorough understanding has yet to be achieved. 
We analyzed the structural deformations of a LaAlO$ _{3} $ (001) slab induced by hydrogen adatoms and oxygen vacancies at its surface by means of density functional theory. Moreover, we investigated the influence of surface reconstruction on the density of states and determined the change of the local density of states at the Fermi level with increasing distance from the surface for bare LaAlO$ _{3} $ and for a conducting LaAlO$_{3}$/SrTiO$_{3}$ interface. In addition, the Al-atom displacements and distortions of the TiO$_6$-octahedra were estimated.

\end{abstract}

\section{Introduction}
\label{intro}
Since the discovery of high temperature superconductivity~\cite{bednorz1986possible} considerable effort has been made to study the behavior of strongly correlated electrons in transition metal oxides. Various types of impurities, crystal structure defects, stoichiometric variations, external electric and magnetic fields, light illuminations, uniaxial or hydrostatic pressures are accessible parameters for the transition metal oxide properties of films and heterostructures.
The structural tuning of films results in a variety of fascinating many-body phenomena.

In particular, the paradigm example of a heterostructure between the non-magnetic and insulating oxides LaAlO$ _{3}$~(LAO) and SrTiO$ _{3}$~(STO)  demonstrates rich physics including the coexistence of superconducting 2D electron liquid~\cite{ohtomo2004high,reyren2007superconducting,thiel2006tunable,yu2014unifying} and magnetism~\cite{yu2014unifying,brinkman2007magnetic,li2011coexistence,wang2011electronic,kalisky2012critical,pavlenko2012oxygen,pavlenko2013}.

The aim of the present study is to investigate the electronic properties and structural distortions of  surfaces and interfaces based on LAO and STO by means of density functional theory.

\section{Method}
We performed \textit{ab-initio} calculations within density functional theory (DFT)~\cite{hohenberg1964inhomogeneous,perdew1996generalized} and the Vienna Ab-Initio Simulation Package (VASP 5.3)~\cite{Kresse1996} implemented into MedeA software~\cite{medea}. Exchange and correlation were included at the level of the generalized gradient approximation (GGA) \cite{perdew1996generalized}. %The Dudarev approach~\cite{dudarev1998electron} was applied within a simplified generalized gradient approximation GGA+U scheme (U$_{Ti}$=2 eV and U$_{La}$=8 eV).
We have used the projector-augmented wave method~\cite{PhysRevB.50.17953,kresse1999ultrasoft} with a plane-wave basis set and a cutoff energy of 400\,eV. The force tolerance was 0.05 eV/\AA~and the energy tolerance for the self-consistency loop was equal to $10^{-5}$\,eV. The Brillouin zone was sampled on a grid of 5$\times$5$\times$1 \textbf{k}-points (7$\times$7$\times$1 for 4 LAO/4.5 STO/4 LAO heterostructure). In order to avoid the interaction of surfaces and slabs with their periodic images during simulation, a 20\,\AA \space vacuum region perpendicular to the surface was added. 

\section{Results and discussions}

\subsection{Surface properties of LAO and STO}
\label{sec:3}

We considered the electronic structure of an LAO (001) slab that consisted of 5.5 unit cells with identical terminations on both sides (La$ ^{+3} $O$ ^{-2} $ or Al$ ^{+3} $O$ _{2}^{-2} $). The central region of 1.5 unit cells was kept fixed during the optimization as in Ref.~\cite{krishnaswamy2014structure}.
 It was shown in Ref.~\cite{janotti2012controlling,xie2011control} that the surface reconstruction affects intrinsic transport properties of the slab. Since the La$ ^{+3} $O$ ^{-2}$ layers have a charge of +1 and Al$ ^{+3} $O$ _{2}^{-2} $ layers have a charge of -1 per unit cell, the unreconstructed (1$ \times $1) LaO-terminated (001) surface lacks one electron, whereas the AlO$ _{2} $-terminating surface has an excess of one electron. We confirmed that the LAO surface with AlO$ _{2}$-termination has lower energy per area and we will therefore focus on this structure in the discussion below. 

\begin{figure}[h!]
	\centering
		\begin{minipage}[h]{0.23\linewidth}
			\center{\includegraphics[width=1\linewidth]{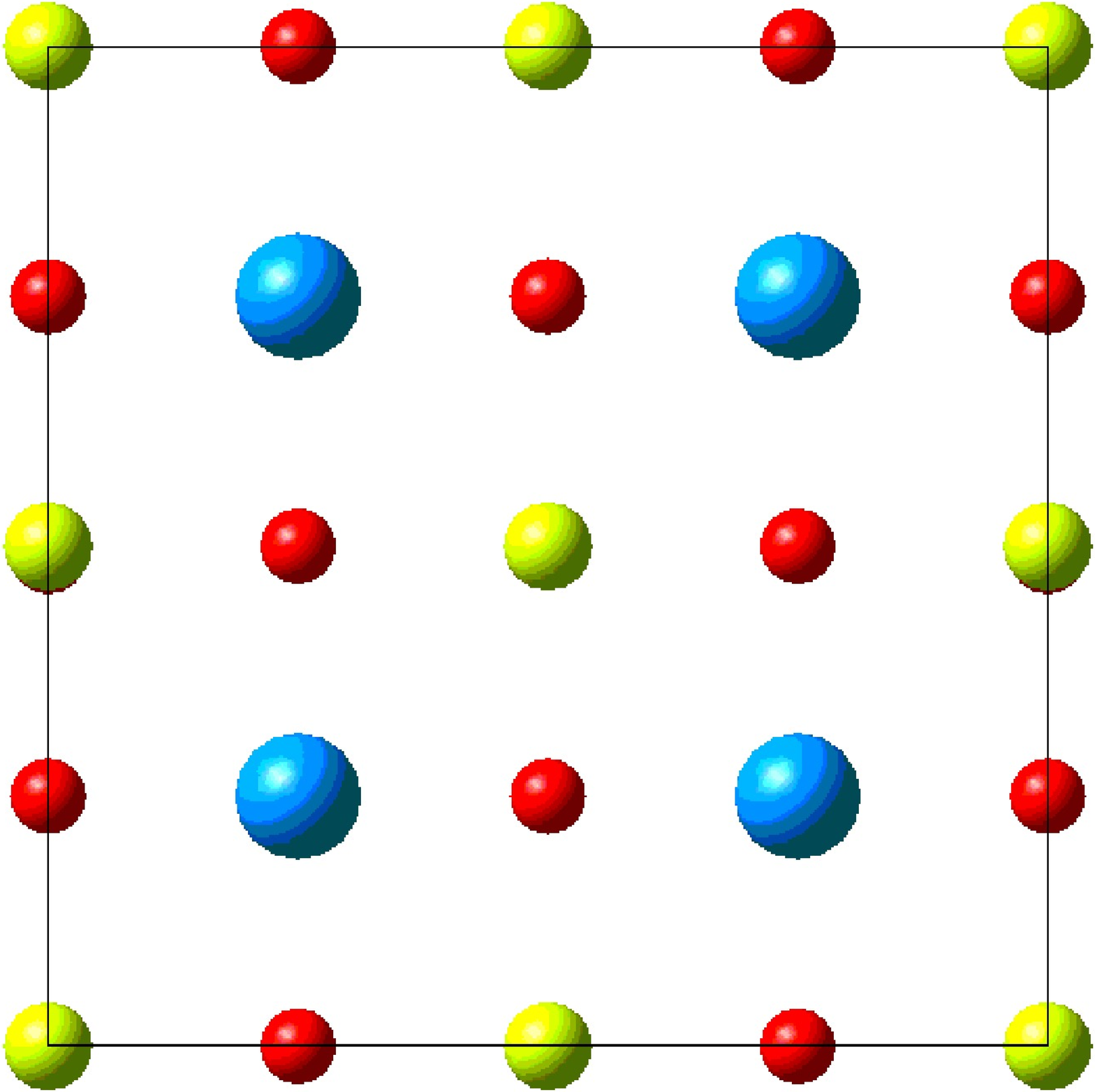}  (a)\\}
		\end{minipage}
			\hfill
	\begin{minipage}[h]{0.33\linewidth}
	\center{\includegraphics[width=1\linewidth]{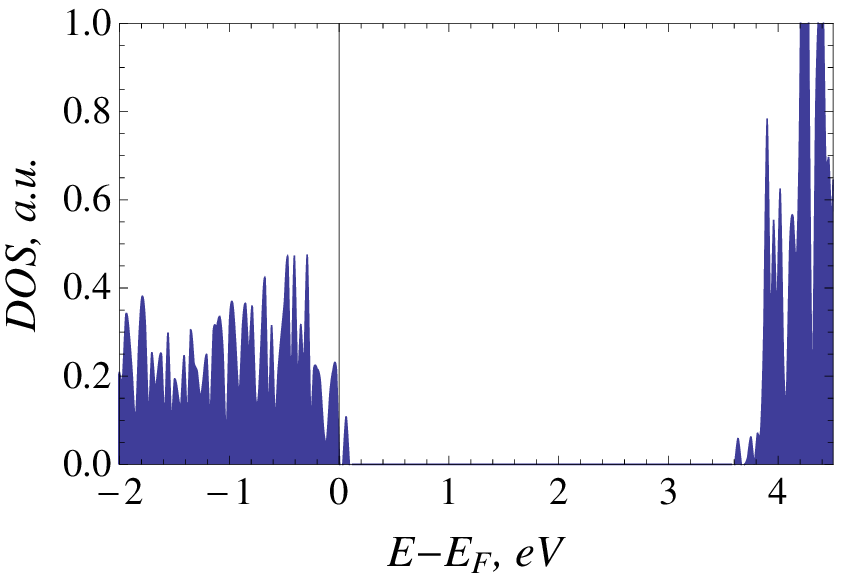}  (b)\\}
	\end{minipage}
\hfill
	\begin{minipage}[h]{0.33\linewidth}
		\center{\includegraphics[width=1\linewidth]{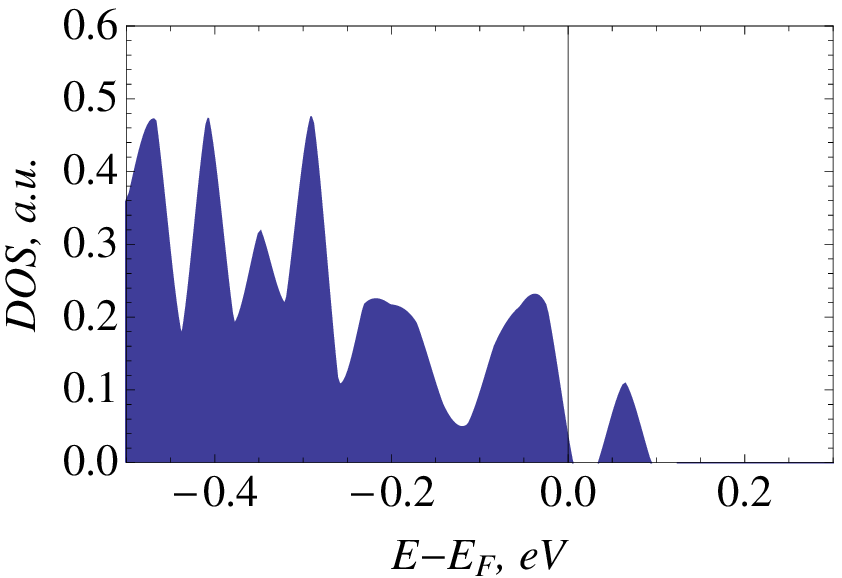} (c) }
	\end{minipage}
	\caption{Top view of an optimized (2$ \times $2) bare LaAlO$_{3}$ (001) surface with AlO$_{2}$-termination (a), corresponding density of states (DOS) profile (b) and its enlarged version near the Fermi level (c). Red spheres denote oxygen, blue spheres denote lanthanum and yellow spheres denote aluminum.}
	\label{ris:dos_lao}
\end{figure}

Under experimental conditions, thin films of LAO are grown on top of the STO substrate. The lattice constant $\textit{a}$ of LAO approaches the in-plane lattice parameter $\textit{a}$ of STO \cite{ohtomo2004high},therefore the LAO lattice parameter was fixed at this value (\textit{a}=3.941\,\AA \space from our GGA calculation for bulk STO). It follows from our findings (see Fig.~\ref{ris:dos_lao}) that the bare LAO surface with AlO$_{2}$-termination is a semiconductor with a very small gap separating the bulk bands and a narrow surface level.

\begin{figure}[h!]
	\centering
	
	\begin{minipage}[h]{0.23\linewidth}
	\center{\includegraphics[width=1\linewidth]{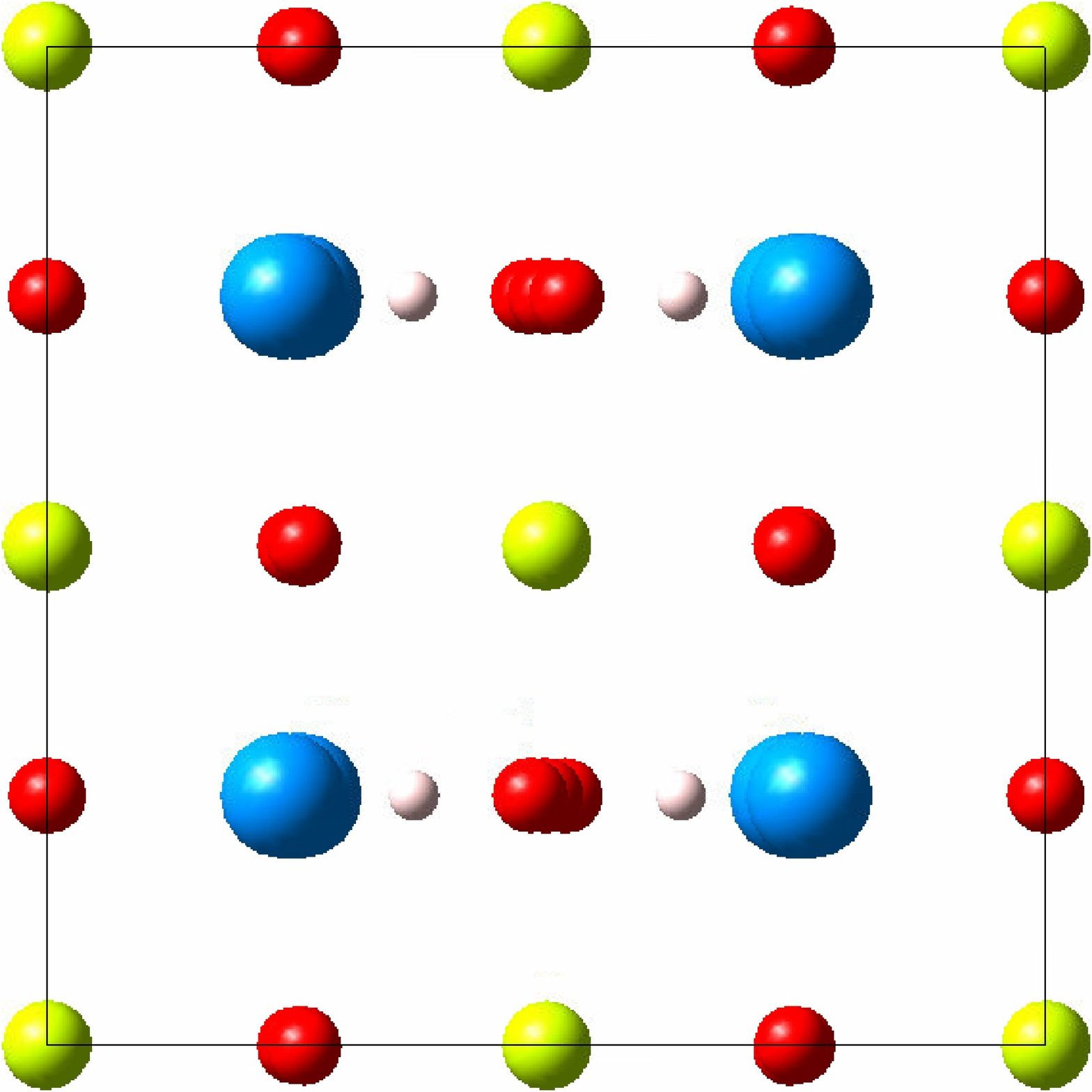}  (a)\\}
	\end{minipage}
	\hfill
		\begin{minipage}[h]{0.33\linewidth}
			\center{\includegraphics[width=1\linewidth]{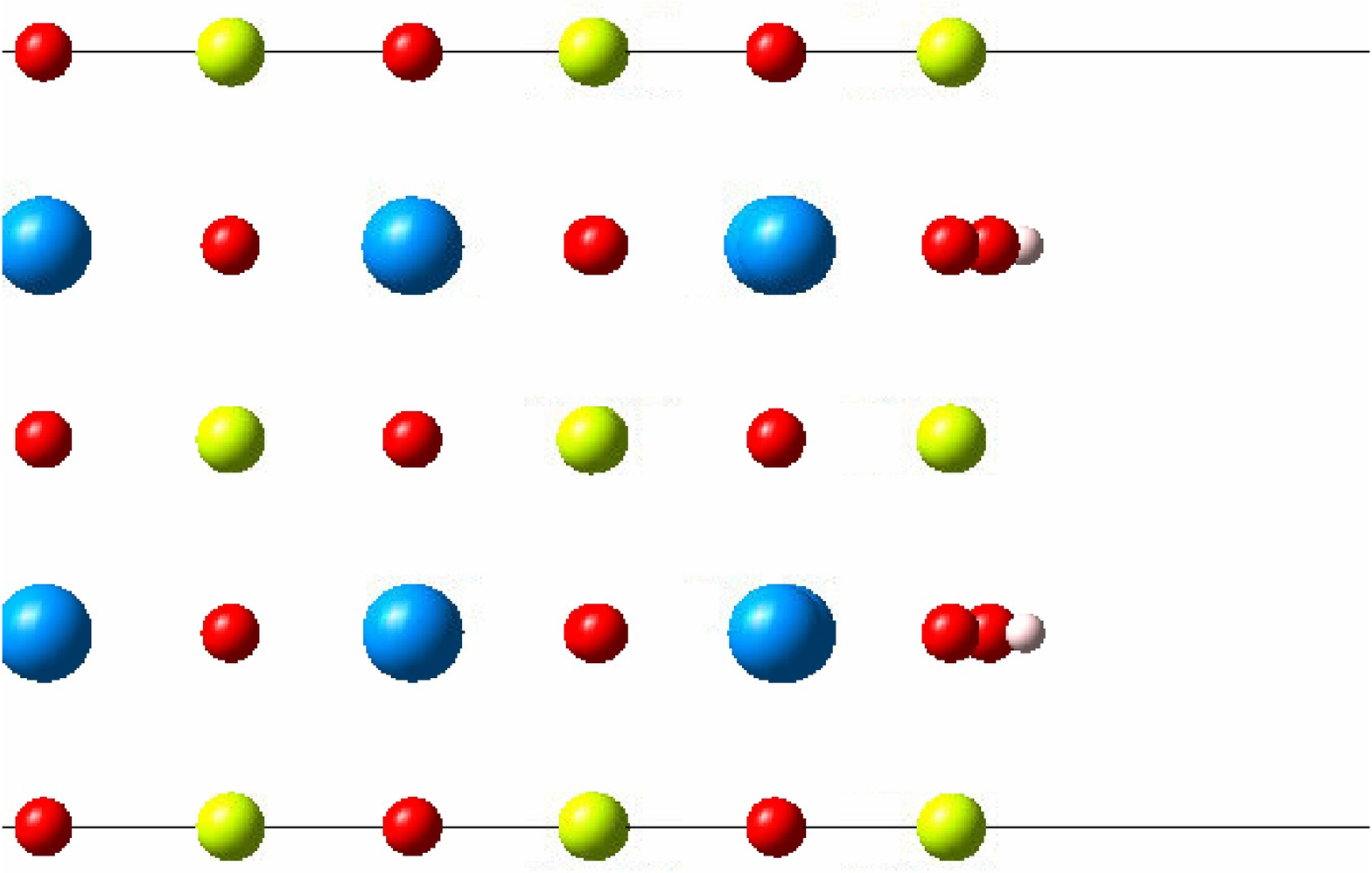}  (b)\\}
		\end{minipage}
	\hfill
	\begin{minipage}[h!]{0.33\linewidth}
	\center{\includegraphics[width=1\linewidth]{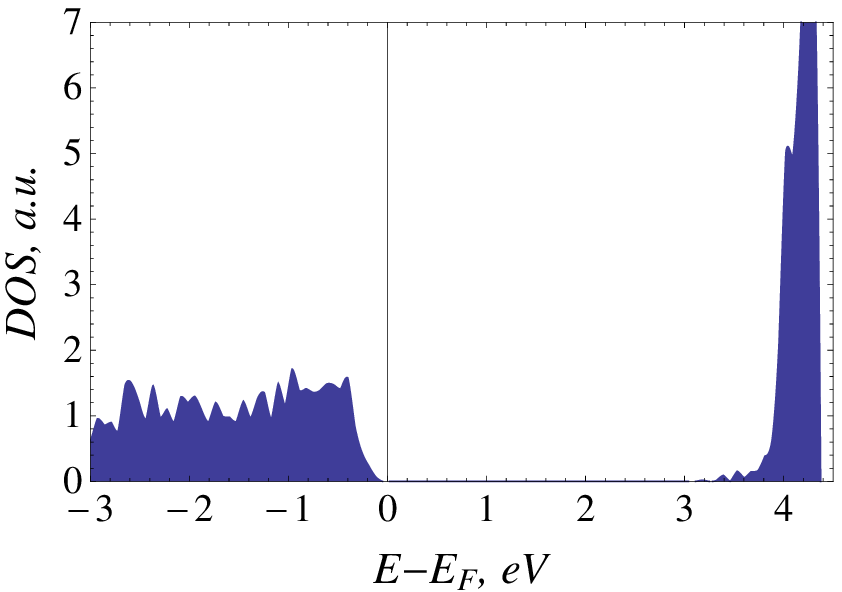} (c)\\ }
	\end{minipage}
	\caption{Top (a) and side (b) view of an optimized (2$ \times $2) LaAlO$_{3}$ (001) surface with AlO$_{2}$-termination and H-adatoms  with corresponding density of states (c). Red spheres denote oxygen, blue spheres denote lanthanum and yellow spheres denote aluminumm, and white spheres denote hydrogen.}
	\label{ris:Hadaton}
\end{figure}

It is known that the presence of impurities on the surface affects the final termination layer of LAO during surface growth~\cite{yao1998thermal}. We considered the case of a hydrogen (H) adatom which is present in almost all growth and annealing environments. The top view of the optimized (2$\times$2) LaAlO$_{3}$ surface with H-adatom at the surface and corresponding DOS are shown in Fig.~\ref{ris:Hadaton}. As in the work of Krishnaswamy \textit{et al.}~\cite{krishnaswamy2014structure} the H-adatom bonds to oxygen. It takes a position between the strontium and oxygen and is located above the surface plane. Comparing  Fig.~\ref{ris:dos_lao} b, c and Fig.~\ref{ris:Hadaton} c it follows that the presence of a H-adatom leads to afilling of the empty states right above the Fermi level of pure LAO and, therefore,a suppression of the albeit small surface conductivity.
\vspace{-15pt}
\vspace{-15pt}
\begin{figure}[h!]
	\centering
		\begin{minipage}[h]{0.23\linewidth}
		\centering
		\center{\includegraphics[width=1\linewidth]{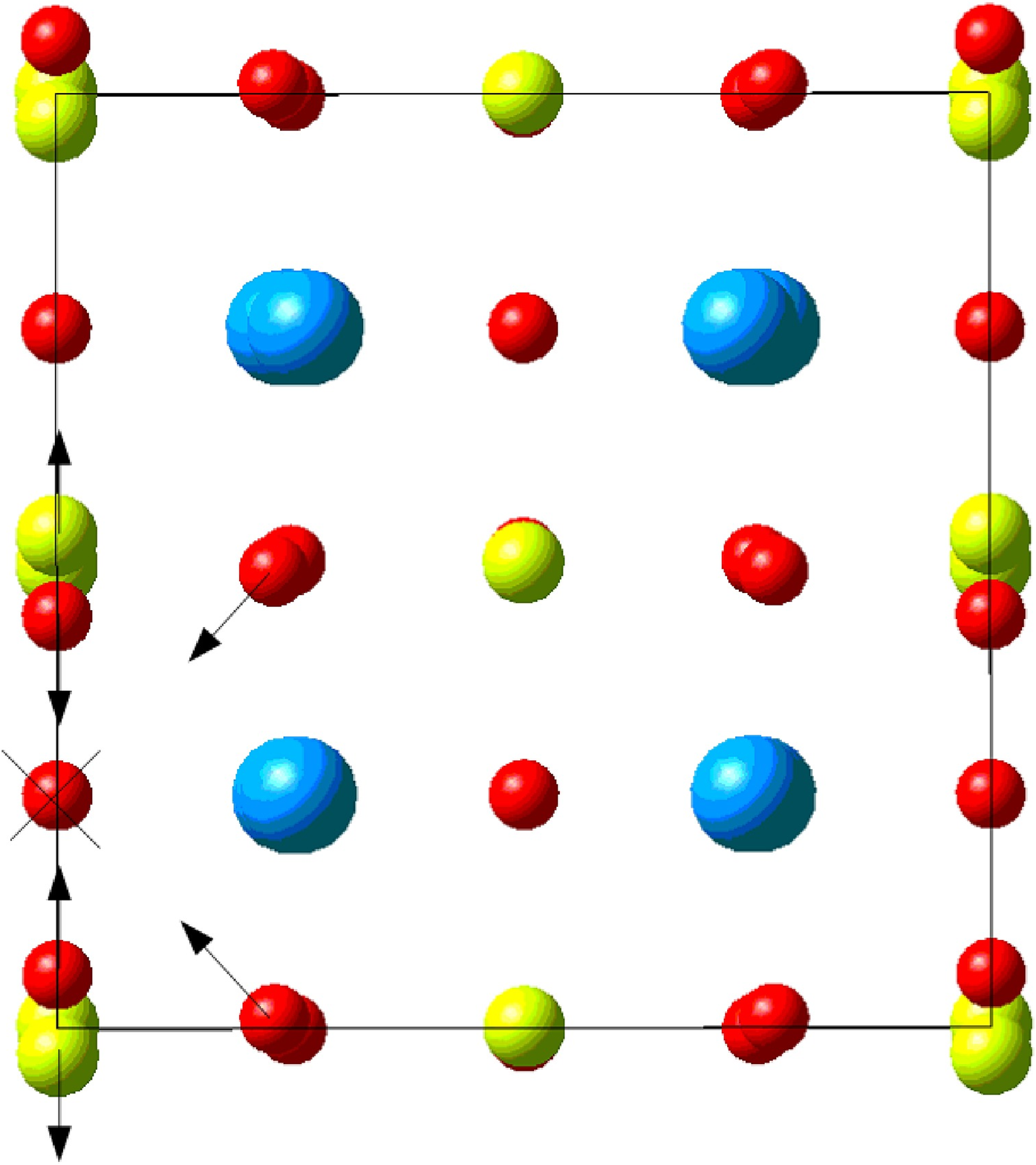}  (a)\\}
	\end{minipage}
		\hfill
	\begin{minipage}[h]{0.33\linewidth}
		\centering
		\center{\includegraphics[width=1\linewidth]{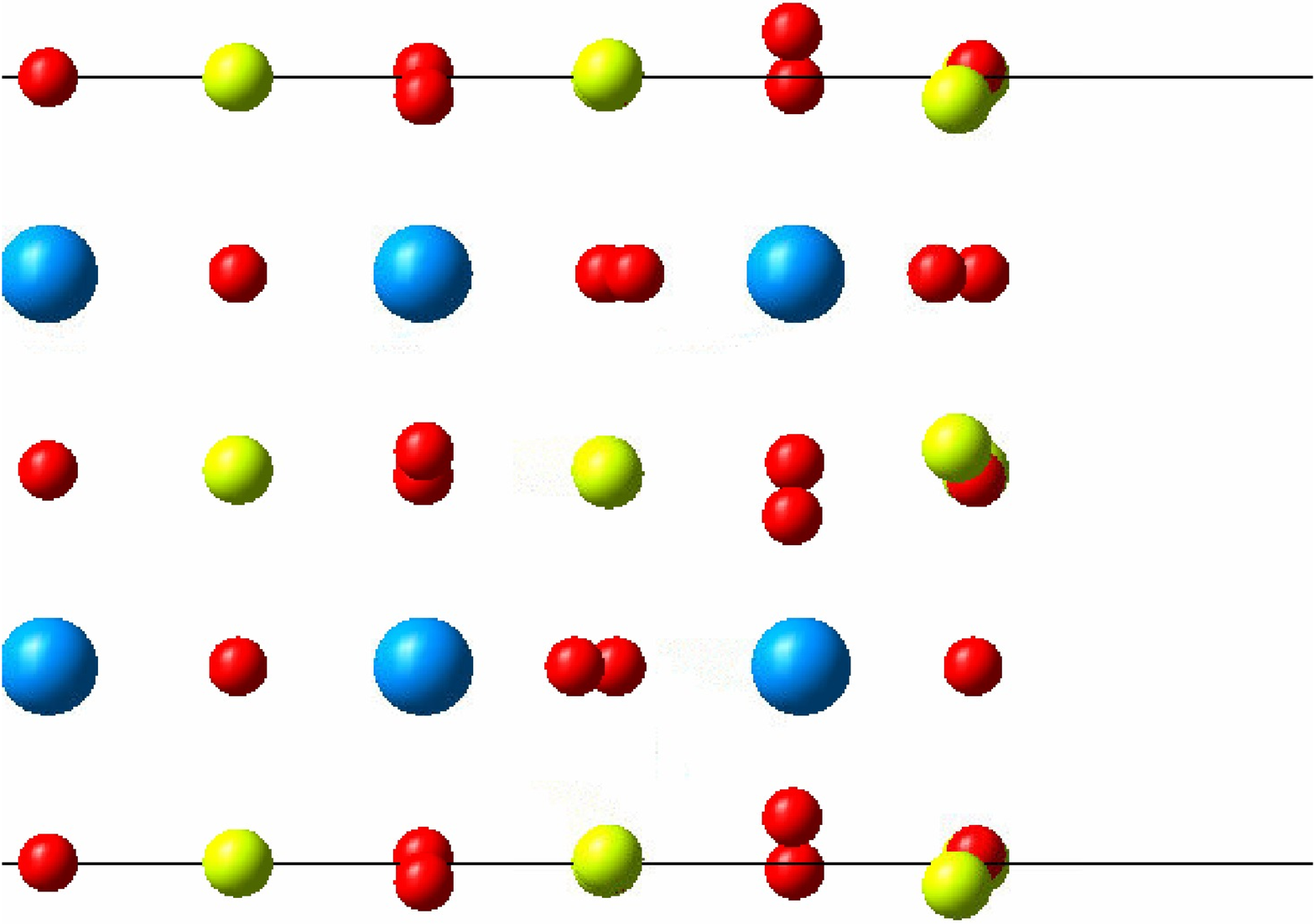}  (b)\\}
	\end{minipage}
		\hfill
	\begin{minipage}[h!]{0.33\linewidth}
		\center{\includegraphics[width=1\linewidth]{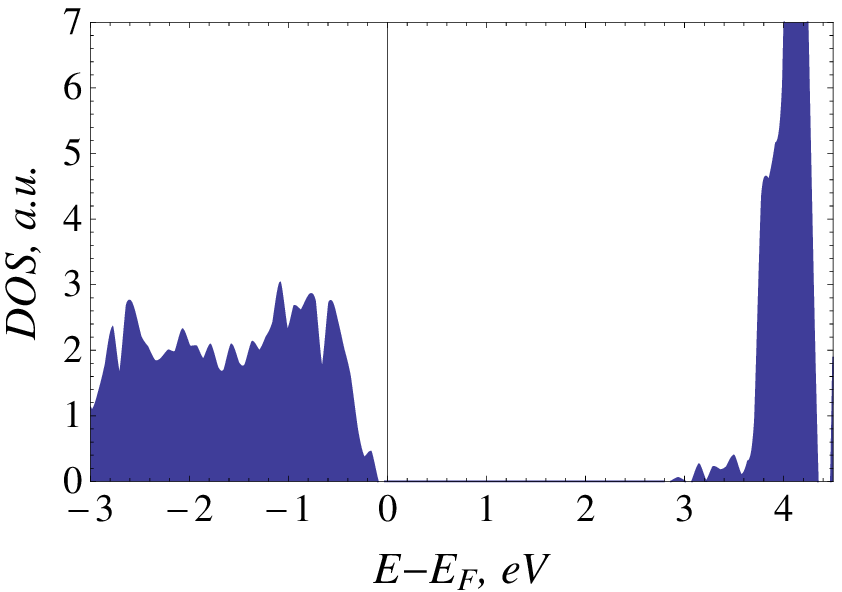} (c)\\ }
	\end{minipage}
	\caption{Top (a) and side (b) view of an optimized (2$ \times $1) LaAlO$_{3}$ (001) surface with AlO$_{2}$-termination and O-vacancy with corresponding density of states (c). Red spheres denote oxygen, blue spheres denote lanthanum, yellow spheres denote aluminum, $ \times $ indicates the oxygen vacancy and arrows point the atoms displacement.}
	\label{ris:O-vac}
\end{figure}

The second type of surface reconstruction is caused by an oxygen (O) vacancy. This type is one of the most studied~\cite{pavlenko2012oxygen,kalabukhov2007effect,vonk2012polar,li2011formation,zhang2010origin}. For this type of reconstruction we took a (2$ \times $2) cell (Fig.~\ref{ris:O-vac}). As in the case of H-adatoms  we obtained a suppression of surface conductivity (Fig.~\ref{ris:O-vac} c). We performed the same calculations for an STO slab and in contrast to LAO, the STO surface did not show empty states just above the Fermi level and exhibited the same insulator behavior as the bulk STO.

\subsection{LAO/STO/LAO heterostructure properties}
\begin{figure} [h!]
\centering
\includegraphics[width=1\linewidth]{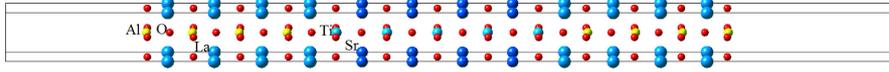}
\caption{Structure of 4 LAO/4.5 STO/4 LAO superlattice.}
\label{fig:LAO_STO_STRUCTURE}
\end{figure}

The heterostructures in our calculations consisted of a central region of SrTiO$ _{3} $ (with fixed number of layers $N_\mathrm{STO}=4.5$) bounded on both sides with a varying number of LaAlO$ _{3} $ layers (see Fig.~\ref{fig:LAO_STO_STRUCTURE}). According to experimental evidence~\cite{ohtomo2004high,thiel2006tunable}, the \textit{n}-type contact with (TiO$ _{2} $)$ ^{0} $-(LaO)$ ^{+} $ shows conductivity at the interface and, therefore, we have focused on this type of interface.
%\vspace{-10pt}
\label{sec:4}
\vspace{-25pt}

 \begin{figure}[h!]
 	\begin{minipage}[h]{0.33\linewidth}
 		\centering
 		\center{\includegraphics[width=1\linewidth]{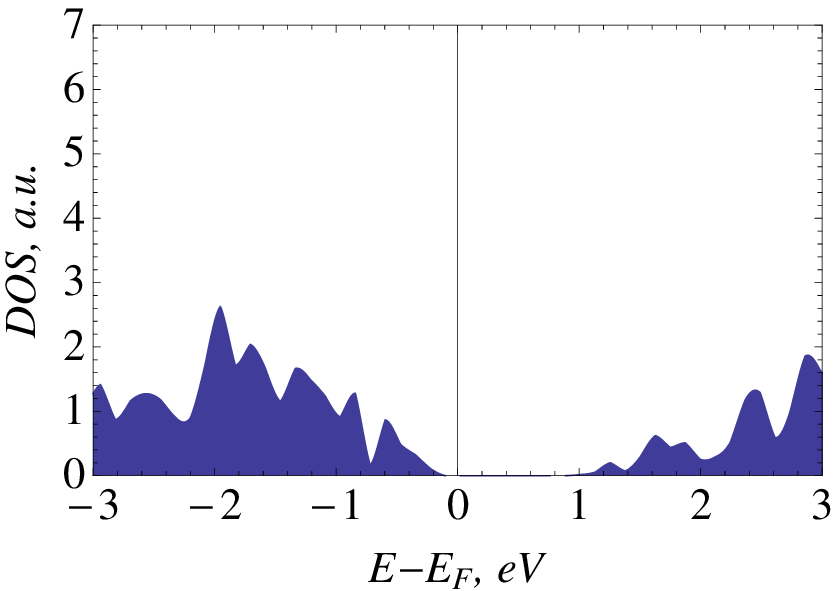}  (a)\\}
 	\end{minipage}
 	%\hfill
 	\begin{minipage}[h!]{0.33\linewidth}
 	\center{\includegraphics[width=1\linewidth]{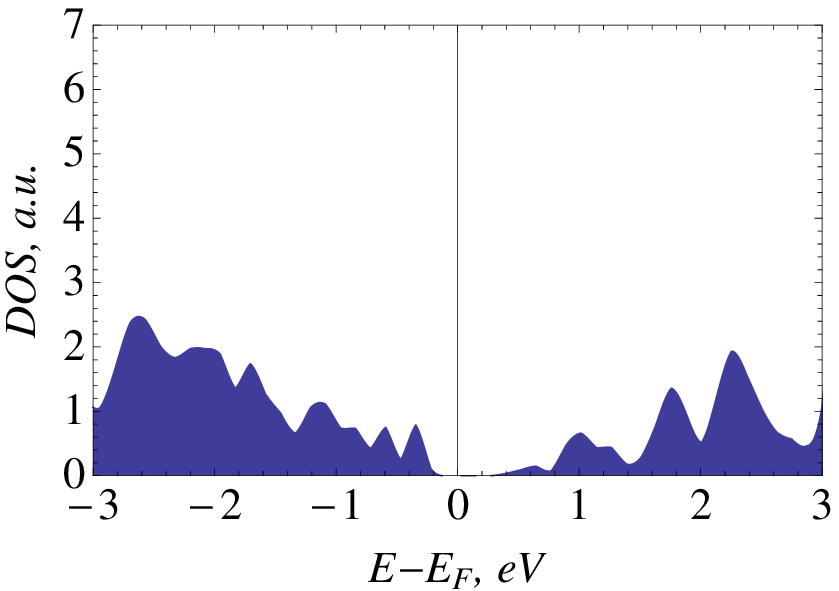} (b)\\ }
 	\end{minipage}
 	\begin{minipage}[h!]{0.33\linewidth}
		\center{\includegraphics[width=1\linewidth]{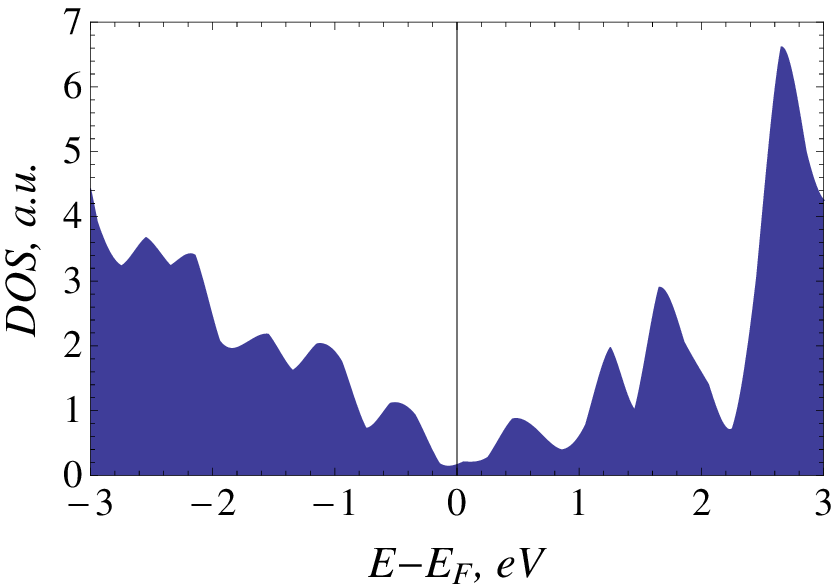} (c)\\ }
 	\end{minipage}
 	\caption{Density of states plots for  LAO/4.5 STO/LAO superlattice interface with (a) 2 LAO layers, (b) 3 LAO layers, (c) 4 LAO layers.}
 	\label{ris:dos_interface}
 \end{figure}
 
\begin{figure}%[h!]{0.9\linewidth}
	\centering
	\includegraphics[width=0.9\linewidth]{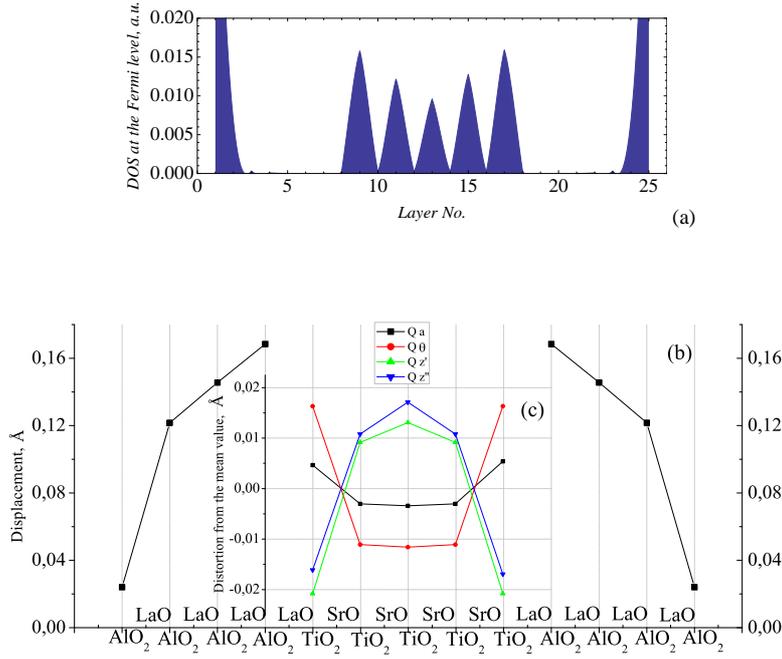} 
	%	\end{figure}
	\caption{Projection onto the different layers of 4 LAO/4.5 STO/4 LAO superlattice:  density of state at the Fermi level calculated in a small energy window (40 meV) (a), displacement of Al atoms with respect to oxygen plane (b) and symmetrized displacements Q (normal coordinates) of TiO$_6$ octahedral systems ~\cite{bersuker2006jahn}(c).}
	\label{layers}
\end{figure}

We calculated the density of states profiles for a varying number of LAO layers (Fig.~\ref{ris:dos_interface} a-c). It was found that the band gap decreases with an increase in the number of LAO layers. At four LAO layers the band gap of the LAO/STO/LAO heterostructure vanishes (in agreement with~\cite{cossu2013metal}). Layer resolved analysis of the density of states and structural deformation was performed for this system (Fig.~\ref{layers}).

Fig.~\ref{layers} a presents the spatially resolved DOS at the Fermi level $\textit{n}$($\textit{E}$$_{F}$) of a 4 LAO/4.5 STO/4 LAO heterostructure. The maximum intensity corresponds to surface AlO$_{2}$ and the TiO$ _{2} $ layers on the interface and agrees well with previous findings~\cite{pentcheva2006charge,pavlenko2012oxygen,ohtomo2004high}. The value of  $\textit{n}$($\textit{E}$$_{F}$) does not vanish in the middle of the STO slab due to its small thickness during simulations. Orbital decomposition of DOS indicates that the electronic states at the Fermi level consist entirely of Ti 3d$ _{xy} $ states, which agrees with previous work~\cite{pavlenko2012oxygen,pavlenko2013,pentcheva2006charge}. As in the case of a bare LAO surface with AlO$ _{2} $ termination, there is a contribution to $\textit{n}$($\textit{E}$$_{F}$) from the surface layers.

\begin{figure}%[h]{0.6\linewidth}
	\centering 
	\includegraphics[width=0.6\linewidth]{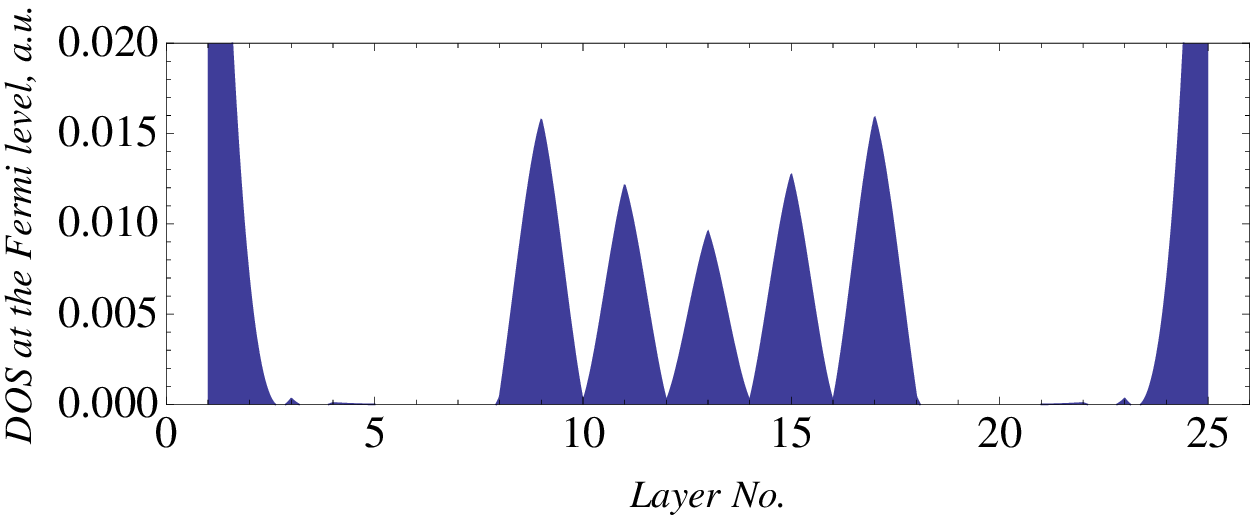} (a)
\end{figure}
%\vspace{-25pt}

As a result of the DFT structural optimization we obtained distortions of the LAO/STO heterostructure, see Fig. \ref{fig:LAO_STO_STRUCTURE}.
Ti atoms moved towards the center of the slab, whereas La atoms shifted towards the surface. Fig.~\ref{layers} b shows the displacement of Al atoms out of the oxygen planes. The biggest shift was obtained for Al atoms near the LAO/STO interface. The symmetrized displacements, $Q$, (normal coordinates) of TiO$_6$ octahedral system are presented in Fig.~\ref{layers} c. $Q_{i} $ indicates the collective atom displacements, which, under the symmetry operation, transform according to one of its irreducible representations~\cite{bersuker2006jahn}. The smallest magnitude was found for Q$ _{a}$ breathing mode, which corresponds to a totally symmetric shifting of O atoms out of the center Ti atom. Further non-zero normal coordinates, $Q$, were associated with oxygen movements in $z$-direction perpendicular to the interface, and they led to the splitting of t$_\mathrm{2g}$ orbital.

\section{Conclusion}
In the present work, by means of DFT calculations on the GGA level of theory, we have investigated the structural and electronic properties of the surfaces and interfaces based on the two insulators, LAO and STO.  We found that the bare surface of LAO with AlO$ _{2} $-termination has conductive surface layers, whereas the STO surface is insulating. Two considered types of surface defects (H-adatom and O-vacancy) of the AlO$ _{2} $-terminated (001) LAO slab lead to a suppression of surface conductivity. 
For the LAO/STO/LAO interface, an insulator-metal transition was found by increasing the number of LAO layers from three to four. The largest contribution to the interface conductivity is from the TiO$ _{2}$ layers (Ti d$_{xy}$ state), decreasing with distance from the interface. We also obtained surface conductivity, which we expect to be suppressed by defects as for the case of the LAO slab.

%The local structural deformations of TiO$_6$ octahedrons and Al atoms displacements were analysed and its decay away from the interface was shown as well.

%In conclusion, our findings obtained within GGA level of theory agree well with previous results obtained with more time consuming %GGA+U~\cite{pavlenko2012oxygen} and HSE~\cite{:/content,cossu2013metal,krishnaswamy2014structure} methods. The chosen computational parameters could be used for further investigations not only of this system, but for others metal-oxide surfaces and interfaces. 

\begin{acknowledgements}
The reported study was supported by the Supercomputing Center of Lomonosov Moscow State University with support of the Russian Government Program of Competitive Growth of Kazan Federal University and of the DFG Sonderforschungsbereich TRR 80. The work of A.G.~Kiiamov  was funded by the subsidy allocated to Kazan Federal University for the state assignment in the sphere of scientific activities. The authors acknowledge helpful discussion with D. Juraschek, A. Petrova, Kate Reidy and Jessica Weitbretch.

\end{acknowledgements}


\begin{thebibliography}{10}
%1
\bibitem{bednorz1986possible}
J.~G.~Bednorz, K.~A.~M\"uller, 
Z.\ Phys.\ B \textbf{64}, 189 (1986). 

%\bibitem{takahashi2001interface} K.~Takahashi, M.~Kawasaki, and Y.~Tokura, Appl.\ Phys.\ Lett.\ \textbf{79}, 1324 (2001). 

%2
\bibitem{ohtomo2004high}
A.~Ohtomo and H.~Hwang, 
Nature \textbf{427}, 423 (2004). 

%3
\bibitem{reyren2007superconducting}
N.~Reyren, S.~Thiel, A.~D.~Caviglia, L.~F.~Kourkoutis, G.~Hammerl, 
C.~Richter, C.~W.~Schneider, T.~Kopp, A.-S.~R\"uetschi, D.~Jaccard, 
M.~Gabay, D.~A.~Muller, J.-M.~Triscone, and J.\ Mannhart, 
Science \textbf{317}, 1196 (2007). 

%4
\bibitem{thiel2006tunable}
S.~Thiel, G.~Hammerl, A.~Schmehl, C.~Schneider, J.~Mannhart, 
Science \textbf{313}, 1942 (2006). 

%5
\bibitem{yu2014unifying}
Y.~Yu and A.~Zunger, 
Nature Commun.\ \textbf{5}, 5118 (2014). 
%6
\bibitem{brinkman2007magnetic}
A.~Brinkman, M.~Huijben, M.~Van~Zalk, J.~Huijben, U.~Zeitler, J.~Maan,
W.~Van~der Wiel, G.~Rijnders, D.~Blank, and H.~Hilgenkamp, 
Nature Mat.\ \textbf{6}, 493 (2007). 
%7
\bibitem{li2011coexistence}
L.~Li, C.~Richter, J.~Mannhart, and R.~Ashoori, 
Nature Phys.\ \textbf{7}, 762 (2011). 
%8
\bibitem{wang2011electronic}
Ariando, X.~Wang, G.~Baskaran, Z.~Q.\ Liu, J.~Huijben, J.~B.\ Yi, 
A.~Annadi, A.\ R.\ Barman, A.~Rusydi, S.~Dhar, Y.~P.\ Feng, J.\ Ding, 
H.\ Hilgenkamp, and T.\ Venkatesan, 
Nature Commun.\ \textbf{2}, 188 (2011). 
%9
\bibitem{kalisky2012critical}
B.~Kalisky, J.\ A.\  Bert, B.\ B.\  Klopfer, C.~Bell, H.\ K.\  Sato, 
M.~Hosoda, Y.~Hikita, H.\ Y.\  Hwang, and K.\ A.\  Moler, 
Nature Commun.\ \textbf{3}, 922 (2012). 
%10
\bibitem{pavlenko2012oxygen}
N.~Pavlenko, T.~Kopp, E.~Tsymbal, J.~Mannhart, and G.~Sawatzky, 
Phys.\ Rev.\ B \textbf{86}, 064431 (2012). 
%11
\bibitem{pavlenko2013}
N.~Pavlenko, T.~Kopp, J.~Mannhart, Phys.\ Rev.\ B \textbf{88} 201104 (2013). 
%12
\bibitem{hohenberg1964inhomogeneous}
P.~Hohenberg and W.~Kohn, 
Phys.\ Rev.\ \textbf{136}, B864 (1964). 
%13
\bibitem{perdew1996generalized}
J.P. Perdew, K.~Burke, and M.~Ernzerhof, 
Phys.\ Rev.\ Lett.\ \textbf{77}, 3865 (1996); W. Kohn and L. J. Sham, Phys.\ Rev.\ A\ \textbf{140}, 1133 (1965)
%14
\bibitem{Kresse1996}
G.~Kresse and J.~Furthm\"uller, 
Phys.\ Rev.\ B \textbf{54}, 169 (1996). 
%15
\bibitem{medea}
MedeA\textsuperscript{\textregistered}-2.17, Materials Design, Inc., 
Angel Fire, NM, USA, 2015. 
%16
\bibitem{PhysRevB.50.17953}
P.\ E.\ Bl\"ochl, Phys.\ Rev.\ B \textbf{50}, 17953 (1994). 
%17
\bibitem{kresse1999ultrasoft}
G.~Kresse and D.~Joubert, Phys.\ Rev.\ B \textbf{59}, 1758 (1999). 
%18
\bibitem{krishnaswamy2014structure}
K.~Krishnaswamy, C.~Dreyer, A.~Janotti, and C.~Van~de Walle, 
Phys.\ Rev.\ B \textbf{90}, 235436 (2014). 
%19
\bibitem{janotti2012controlling}
A.~Janotti, L.~Bjaalie, L.~Gordon, and C.~Van~de Walle, 
Phys.\ Rev.\ B \textbf{86}, 241108 (2012). 
%20
\bibitem{xie2011control}
Y.~Xie, Y.~Hikita, C.~Bell, H.\ Y.\ Hwang, 
Nature Commun.\ \textbf{2}, 494 (2011). 
%21
\bibitem{yao1998thermal}
J.~Yao, P.~Merrill, S.~Perry, D.~Marton, and J.~Rabalais, 
J.\ Chem.\ Phys.\ \textbf{108}, 1645 (1998). 
%22
\bibitem{kalabukhov2007effect}
A.~Kalabukhov, R.~Gunnarsson, J.~B\"orjesson, E.~Olsson, T.~Claeson, and 
D.~Winkler, 
Phys.\ Rev.\ B \textbf{75}, 121404 (2007). 
%23
\bibitem{vonk2012polar}
V.~Vonk, J.~Huijben, D.~Kukuruznyak, A.~Stierle, H.~Hilgenkamp, 
A.~Brinkman, and S.~Harkema, 
Phys.\ Rev.\ B \textbf{85}, 045401 (2012).
%24
\bibitem{li2011formation}
Y.~Li, S.N. Phattalung, S.~Limpijumnong, J.~Kim, and J.~Yu, 
Phys.\ Review B \textbf{84}, 245307 (2011). 
%25
\bibitem{zhang2010origin}
L.~Zhang, X.F. Zhou, H.T. Wang, J.J. Xu, J.~Li, E.~Wang, and S.\ H.\ Wei, 
Phys.\ Rev.\ B \textbf{82}, 125412 (2010). 
%26
\bibitem{cossu2013metal}
F.~Cossu, U.~Schwingenschl\"ogl, and V.~Eyert, 
Phys.\ Rev.\ B \textbf{88}, 045119 (2013)
%27
\bibitem{pentcheva2006charge}
R.~Pentcheva and W.\ E.\ Pickett, 
Phys.\ Rev.\ B \textbf{74}, 035112 (2006). 
%28
\bibitem{bersuker2006jahn}
I.\ B.\ Bersuker, 
{\em The Jahn-Teller Effect}, 
Cambridge University Press, (2006). 

.
	

\end{thebibliography}
\end{document}